# Relating Eye-Tracking Measures With Changes In Knowledge on Search Tasks

Nilavra Bhattacharya
School of Information
University of Texas, Austin
nilavra@ieee.org

Jacek Gwizdka
School of Information
University of Texas, Austin
etra2018@gwizdka.com

## ABSTRACT

We conducted an eye-tracking study where 30 participants performed searches on the web. We measured their topical knowledge before and after each task. Their eye-fixations were labelled as "reading" or "scanning". The series of reading fixations in a line, called "reading-sequences" were characterized by their length in pixels, fixation duration, and the number of fixations making up the sequence. We hypothesize that differences in knowledge-change of participants are reflected in their eye-tracking measures related to reading. Our results show that the participants with higher change in knowledge differ significantly in terms of their total reading-sequence-length, reading-sequence-duration, and number of reading fixations, when compared to participants with lower knowledge-change.

## 1 INTRODUCTION AND BACKGROUND

Online search is a ubiquitous activity. As noted by [Brookes 1980] in the "fundamental equation" of information and knowledge (p. 131), users perform searches to satisfy their information needs, and gain knowledge in the process. The newly created sub-field 'search-as-learning' (SAL) focuses on studying users' change in knowledge, while and as a result of performing search tasks.

Our investigation is focused on changes in person's declarative knowledge. Model of learning outcomes (CSALO) [Kraiger et al. 1993] posits that learning outcomes are reflected in changes in verbal knowledge, knowledge organization, and cognitive strategies.

In this study, we are interested in changes in verbal knowledge and how it is reflected in eye-tracking (ET) based measures.

The eye-mind link hypothesis [Just and Carpenter 1987] makes ET attractive for assessing acquisition of visually presented information from external environments by humans. Not surprisingly many researchers investigated ET measures in such contexts. For example, psychologists who study reading have found that fixation time increases with difficult and more infrequent words [Pollatsek et al. 2008; Raney and Rayner 1995; Rayner 1998]. Many researchers have focused on studying eye-movement patterns that reflect differences in knowledge and expertise levels. For instance, [Eivazi et al. 2012] saw that while watching videos of neurosurgery, expert surgeons have scan-paths that are more compact and locally defined, and their focus of attention changes less often than novice surgeons. When asked to detect errors and predict the output of source codes, [Nivala et al. 2016] noted that advanced programmers employ shorter fixations and saccades than new programmers, especially when predicting the output. Football coaches were found to have different scan-paths than coaching novices (but not playing novices), when watching a clip from a game video [Iwatsuki et al. 2016]. Total fixation duration, reading speed, and reading length were found to have power to differentiate between users with high and low domain knowledge [Cole et al. 2013]. In general, users with more knowledge and expertise tend to have their fixations (and possibly attention) restricted to areas which are critical, or more relevant to solving the task at hand [Jarodzka et al. 2010; Kardan and Conati 2012], while the fixations of novices tend to wander in non-relevant areas as well. However, to the best of our knowledge, no studies have yet been done to relate ET measures with the learning or knowledge-gain process (i.e. the change in expertise) that occurs during online search.

In this paper, we investigate whether a searcher's ET measures reflect changes in his / her knowledge, without knowing the context of the search task, or attributes of the searcher. In particular, we examined the associations between ET measures, and changes in knowledge, in multi-aspectual search tasks on health-related topics. We conducted a lab-based ET study and asked the following general research question:

*RQ: Are the changes in verbal knowledge, from before to after a search task, observable in eye-tracking (ET) measures?*

## 2 METHOD

### 2.1 Experimental Design

A controlled experiment was conducted in the Information eXperience lab at University of Texas at Austin ($N = 30$), in which voluntary participants were asked to find health related information on the internet. They were pre-screened for English native

**Table 1: Task Prompts**

> *Tasks A1: Vitamin A* – Your teenage cousin has asked your advice in regard to taking vitamin A for health improvement purposes. You have heard conflicting reports about the effects of vitamin A, and you want to explore this topic to help your cousin. Specifically, you want to know:
> (1) *What is the recommended dosage of vitamin A for underweight teenagers?*
> (2) *What are the health benefits of taking vitamin A? Please find at least 3 benefits and 3 disadvantages of vitamin A.*
> (3) *What are the consequences of vitamin A deficiency or excess? Please find 3 consequences of vitamin A deficiency and 3 consequences of its excess.*
> (4) *Please find at least 3 food items that are considered as good sources of vitamin A.*
>
> *Tasks A2: Hypotension* – Your friend has hypotension. You are curious about this issue and want to investigate more. Specifically, you want to know:
> (1) *What are the causes of hypotension?*
> (2) *What are the consequences of hypotension?*
> (3) *What are the differences between hypotension and hypertension in terms of symptoms? Please find at least 3 differences in symptoms between them.*
> (4) *What are some medical treatments for hypotension? Which solution would you recommend to your friend if he/she also has a heart condition? Why?*
>
> ***Example 'S' task:*** *Crohn's disease: I know someone who was recently diagnosed, and am curious about the disease.*

level, uncorrected 20/20 vision, and non-expert topic familiarity. All participants reported daily Internet use longer than an hour, and everyday Google usage. Most of them had been searching online for 7 years or more. The majority of participants considered themselves as proficient in online information searches. Due to technical difficulties during the study, data for 4 participants had to be discarded. Usable data is available for 26 participants (16 females; mean participant age 24.5 years). Eye-tracker used was Tobii TX-300 (Tobii Technology AB, Sweden).

## 2.2 Tasks

Each participant performed three search tasks – two assigned (A) tasks, and one self-generated (S) task – in counterbalanced order, with six rotations. The two 'A' tasks simulated a work-task approach which triggers realistic information need for participants [Borlund 2003]. The task prompts are given in Table 1. Search tasks were performed using Internet Explorer, with a custom sidebar on the left. The sidebar displayed (on demand) the current search task description on top, and had bookmarking and note-taking sections below. Bookmarks allowed the participants to save the URLs of webpages, while notes allowed participants to type and / or copy-paste relevant textual descriptions from webpages. Participants were instructed to bookmark a page if they considered it RELEVANT to their task, and to add notes to the page. All bookmarked URLs and entered notes were saved. A training task allowed participants to familiarize themselves with the interface and study procedure.

## 2.3 Procedure

Each session started with assessment of the participant's working memory capacity (WMC) by using memory span task from a cognitive psychology textbook [Francis et al. 2008]. Before each task, participants completed a Pre-Task knowledge assessment which gauged their existing knowledge of the A task. During the task, participants searched public websites using Google, and were asked to bookmark relevant web pages, and take notes of relevant information they found on a page. Each visited webpage was classified as a 'SERP' (Search Engine Result Page) or a 'CONTENT' page, based on whether it was a Google SERP, or any other page, respectively. A CONTENT page was further marked as 'RELEVANT', if the participant bookmarked the page. During the experiment, the participants' eye-gaze, and all other computer-interactions were recorded. At the end of each task, participants completed a Post-Task knowledge assessment. A session typically lasted from 1.5 to 2 hours. On completion, each participant received $25.

In the Pre- and Post-Task knowledge assessments, participants were asked to free-recall as many words or phrases on the topic of the task as they could, without time limit. Since we are interested in determining changes in knowledge of a participant, we have focused on the CONTENT pages, as we assume that most learning (i.e. new knowledge) comes from these pages. We also argue that new knowledge is most likely obtained by reading, and interacting (in the form of note-taking and bookmarking) with CONTENT pages which are perceived as RELEVANT to a task. Preliminary analysis of the user-interaction data was reported in [Gwizdka and Chen 2016]. The ET measures of reading behaviour on RELEVANT CONTENT pages presented in this paper have not been previously reported.

## 2.4 Measures

*2.4.1 Eye-tracking (ET) Measures:* Sighted people acquire information mainly through vision (reading, watching etc.), and the manner of 'seeing' influences their learning process. Our search-tasks were designed to study finding, reading and processing of textual information. So, we used ET measures that reflect reading.

**Table 2: Definitions of Eye-tracking (ET) Measures Per-task**

| | |
|---|---|
| *Rseq_N* | number of reading-sequences |
| *Rseq_px_tot* | total length of the (mostly horizontal) scan-paths obtained by joining the 'reading' fixation points (in pixels) |
| *Rseq_fixn_ct_tot* | total count of 'reading' fixations making up a sequence |
| *Rseq_dur_tot* | total duration of all fixations comprising reading-sequences (in ms) |
| *Reg_N* | total count of backward-regressions |
| *Reg_px_tot* | total length of regressions (in pixels) |

Informed by the E-Z Reader model [Rayner et al. 2011], we assume that: (a) reading is serial, and words are processed one at a time in the order of appearance in text, (b) multiple words can be read in a fixation, as next word can be seen in parafoveal view, and (c) a minimum fixation time (110ms in our implementation) is required for acquiring a word's meaning. A simple, line-oriented classifier [Gwizdka 2014] was used to label fixations as *reading* or *scanning*, by estimating the lengths of pixels in foveal view (2° of visual field [Strasburger et al. 2011] ≈ 100 pixels), and parafoveal view (5° of visual field [Swanson and Fish 1995] ≈ 320 pixels, to the left of the foveal view). Reading fixations formed *reading-sequences*, which represent reading in one line, while scanning fixations were isolated fixations not in the same line. We also calculated eye regressions by considering location of fixations. While in a reading sequence, if the next fixation was located to the left of the current fixation, at a distance corresponding to more than the foveal radius (60 pixels), we considered it a single eye regression. Since we were interested in how much people read, we operationalized the amount of reading in six ways, as described in Table 2.

Calculating the length of reading-sequences in pixels gives a good estimate of the actual reading scan path length, even when the participant scrolled a web page. This is because: (a) people are less likely to scroll during reading, and (b) our labelling algorithm considers estimated size of foveal and parafoveal areas in screen dimensions and also size of typical text lines on screen. The six ways of operationalizing amount of reading constituted six eye-fixation metrics. They were calculated separately for each task, across all RELEVANT CONTENT pages visited by the participant. This is because we assume that participants learn the most from reading CONTENT web pages that are relevant to a task.

*2.4.2 Knowledge-Change (KC) Measures:* Change in knowledge can be measured in several ways. One method is to simply ask participants to rate their topic-familiarity before and after each task [Liu et al. 2013]. The disadvantage is bias, and low accuracy of such measurement. Another approach is to construct questions to test comprehension [Gadiraju et al. 2018], or use Sentence Verification Technique [Freund et al. 2016]. The disadvantage is the need to create task-specific questions.

To avoid the such drawbacks, we considered the entries generated by participants in a free-recall knowledge assessment [Wilson and Wilson 2013]. We argue that during an information search task, if users encounter the topically relevant words which are new to them, then they will be able to reproduce (at least some of) these words during a free recall assessment [Cooper and Pantle 1967; Feigenbaum 1961; Peterson and Peterson 1959; Taki and Khazaei 2011]. We counted the entries in the Pre- and Post-task responses, and analyzed the rank of the nouns used in them. The higher the value of a word rank, the less frequently it is used in common language expressions. Participants' entries could be in a form of keywords or phrases. More entries on a topic are associated with higher topical knowledge. Word analysis was focused on nouns, since nouns are important for carrying meaning. Topical knowledge and expertise gain is typically associated with the use of more sophisticated vocabulary, and is expressed by using words which are more specialized, and occur less frequently in common usage [Jarodzka et al. 2010]. Our two KC measures were calculated as follows:

- Relative difference in the number of items entered after each task, compared to how many were entered before:

$$rel\_change\_in\_items = \frac{items_{post} - items_{pre}}{items_{post}} \quad (1)$$

- Mean-rank of nouns entered after a task:

$$mean\_rank\_POST\_nouns = \frac{\sum_{i=1}^{n} rank_i}{n} \quad (2)$$

We used word-ranks of approximately 1/3 million most frequent English words, taken from Google's Web Trillion Word Corpus [Franz and Brants 2006], as described in chapter 14 of [Segaran and Hammerbacher 2009].

*2.4.3 Independent and Dependent Variables:* In the context of our research questions, ET measures are our dependent variables. The two KC measures constructed from Pre- and Post-task responses are our independent variables. We calculated these variables separately for each task. This is because task-topics could interact with participants' knowledge, and also because A and S tasks could motivate participants' cognition differently. Thus, a 'participant-task' pair was our unit of analysis. We split these participant-tasks into two groups, Lo and Hi, based on their median scores on each of the KC measures. Our overall expectation was that people who read more on RELEVANT CONTENT web pages would gain more topical vocabulary knowledge.

## 3 RESULTS

As explained in Sec. 2.2, the results discussed here are obtained only from reading RELEVANT CONTENT pages visited by the participants for each task. Fig. 1 shows the range of differences of the two independent KC measures for both Hi and Lo groups, and across the three search tasks. We checked the associations between the

**Table 3: Knowledge-Change (KC) Measures.**

| ET measures | Hi & Lo groups | rel_change in_items | | mean_rank POST_nouns | |
|---|---|---|---|---|---|
| | | Task type | | | |
| | | A | S | A | S |
| Rseq_N | Lo: Mean (SD) | 182 (61.58) | 109.63 (38.3) | 188.04 (58.14) | 150.08 (115.97) |
| | Hi: Mean (SD) | 165.83 (68.21) | 127.69 (116.51) | 161.91 (70.22) | 91.81 (27.84) |
| | M-W \|z\| | 0.94 | 0.58 | 1.63 | 2.06* |
| Rseq_px tot | Lo: Mean (SD) | 107419.52 (50212.48) | 68536.75 (26975.84) | 105647.83 (54055.08) | 79264.92 (59259.52) |
| | Hi: Mean (SD) | 80297.61 (37996.21) | 62968.71 (64228.43) | 83415.06 (35642.09) | 53808.04 (36411.74) |
| | M-W \|z\| | 2.00* | 1.59 | 1.22 | 1.54 |
| Rseq_fixn ct_tot | Lo: Mean (SD) | 1160.56 (476.45) | 749.9 (224.42) | 1220.95 (527) | 885.91 (558.37) |
| | Hi: Mean (SD) | 967.41 (449.67) | 680.23 (603.8) | 916.83 (363.42) | 553 (269.34) |
| | M-W \|z\| | 1.68^ | 1.54 | 1.98* | 2.22* |
| Rseq_dur tot | Lo: Mean (SD) | 279015.84 (117358.01) | 185782 (61612.37) | 307400 (151554.77) | 220327.41 (135154.49) |
| | Hi: Mean (SD) | 243294.45 (143780.67) | 158768.46 (143004.64) | 215636.5 (91442.69) | 123690.81 (53240.78) |
| | M-W \|z\| | 1.64 | 1.94^ | 2.27* | 2.65** |
| Reg_N | Lo: Mean (SD) | 204.08 (96.36) | 118.63 (29.39) | 185 (89.37) | 161.08 (152.69) |
| | Hi: Mean (SD) | 151.83 (72.25) | 127.53 (157.95) | 174.2 (90.48) | 88.81 (41.12) |
| | M-W \|z\| | 1.91^ | 1.51 | 0.46 | 1.75^ |
| Reg_px tot | Lo: Mean (SD) | 43274.25 (20594.55) | 27869.34 (12556.54) | 41337.15 (22192.69) | 32844.51 (26943.17) |
| | Hi: Mean (SD) | 31572.26 (15187.38) | 26298.42 (29002.67) | 34150.35 (14917.7) | 22182.89 (16600.57) |
| | M-W \|z\| | 1.96* | 1.48 | 0.91 | 1.66^ |

For Mann Whitney (M-W) statistics, (**) indicates $p < .01$, (*) indicates $p < .05$, and ( ˆ ) indicates approaching .05 significance ($.05 \leq p < .1$).

KC of individual participants on the three tasks (using Spearman's rank correlation and $\chi^2$), and found no significant relationships. Thus, participants tended to score differently on KC measures on different tasks, and our assumption that people perform differently on different tasks holds.

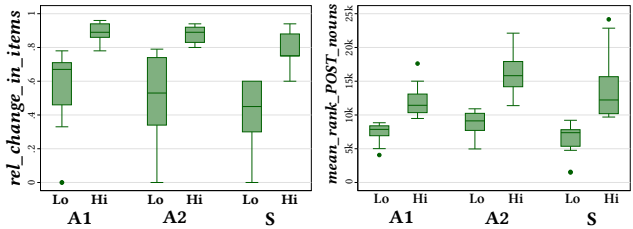

**Figure 1: Range of difference of the two knowledge-change (KC) measures, across task types.**

Table 3 reports values of the ET measures (dependent) for both Hi and Lo groups, for the two task types. Since the reading-sequence measures were not normally distributed, we used non-parametric Mann Whitney U tests to compare the groups. The test statistics are reported in the 'M-W' column of Table 3, and the significant results are marked. Due to the exploratory nature of our study, we conducted M-W tests separately for each ET variable. Our results show that there are significant differences in ET measures for A, and more significant differences for S tasks.

### 3.1 Assigned (A) Tasks

On grouping by *rel_change_in_items*, the total length of all reading-sequences in pixels (*Rseq_px_tot*) was found to differ significantly (Table 3). The Hi group, which entered relatively more phrases after the task than before, had a total reading-sequence length of 80k pixels, while the Lo group had a total reading-sequence length of 107k pixels on average. Significant difference was observed in the total length of regressions (*Reg_px_tot*), with the Hi group regressing 31.5k pixels, and the Lo group regressing 43.2k pixels, per participant. Count of eye regressions was also approaching significant difference ($p = .56$).

When grouping by *mean_rank_POST_nouns*, the total count of fixations in reading-sequences (*Rseq_fixn_ct_tot*), and the total duration of reading-sequences (*Rseq_dur_tot*) were found to differ significantly (Table 3) between the two groups. The Hi group, which entered higher-ranked nouns after the task had a total reading-sequence duration of 215 seconds, and 916 total reading-fixations; whereas the Lo group had a total reading-sequence duration of 307 seconds, and 1220 total reading-fixations, per participant.

### 3.2 Self-generated (S) Tasks

S tasks showed more differences of even higher significance between Lo and Hi groups, than for A tasks. The Hi group, which entered higher-ranked nouns after the task (KC measure: *mean_rank POST_nouns*), had significantly lower values of number of reading-sequences (91 vs. 150), reading-sequence fixation-count (553 vs. 885 fixations) and total fixation duration (123 seconds vs. 220 seconds), when compared to the Lo group. Also, both the eye regression measures were approaching significant difference at 5% level. For KC measure *rel_change_in_items*, the Hi group had a total reading-sequence duration of 158 seconds, while the Lo group had 185 seconds, which was also approaching significant difference.

## 4 DISCUSSION

The mean values of the ET measures for the groups (differing significantly or not) show that the differences are in the opposite direction from our initial expectations. In all cases, the Lo groups have greater values than the Hi groups, indicating that the participants who scored 'higher' on our KC measures did 'less' reading. In general, the total count of fixations in reading-sequences (*Rseq_fixn_ct_tot*), and the total duration of reading-sequences (*Rseq_dur_tot*) were significantly different between Lo and Hi (KC) groups for both tasks. For the A tasks alone, the total lengths of regressions sequences (*Rseq_px_tot*) differed significantly as well.

To understand the above surprising result, we used participants' working memory capacity (WMC) scores. A higher WMC is often associated with better performance and higher intelligence [Colom et al. 2005; Conway et al. 2003]. However, we did not find a significant relationship between our measures of KC and WMC. Therefore, the difference in reading and knowledge test performance do not seem to be able to be explained by the differences in WMC.

Eye regressions differed between the Lo and Hi groups. Where differences were found, the Lo groups moved their eyes backward in reading sequences more often (*Reg_N*) and by a longer distance (*Reg_px_tot*) than the Hi groups. Thus, as reflected in other ET measures, the Lo groups "worked" more on the tasks, yet our KC measures indicate that they learned less. This is a likely indication that Lo groups had more difficulty in acquiring information, and that in spite of investing more effort, they learned less. This could be due to individual differences other than WMC. Data from this study does not allow us to offer a deeper explanation.

## 5 CONCLUSION

In this short paper, we demonstrated significant differences in reading-related eye-movement behaviour between people who learned more, and people who learned less, on online search-tasks. This can be potentially applied to areas where implicit assessment of learning is useful, e.g. online learning environments. For example, if a system equipped with an eye-tracker can detect differences in ET measures between students' behaviour on different lessons, or between different students, the system can offer an intervention for students who are inferred to learn less. Pre-screening ensured that our participants were not topical experts for the assigned (A) search-tasks. However, topics of self-generated (S) tasks were unforeseen. Limitations of our study include (a) using a small number of similar tasks, (b) choice of knowledge change measures, and (c) data analysis at the task level. In future, we plan to add different knowledge-change measures to better reflect the learning process.

## ACKNOWLEDGMENTS

This project is partly funded by IMLS Award #RE-04-11-0062-11 to Jacek Gwizdka.